\documentclass[12pt]{iopart}

\usepackage{times}

\begin{document}

\bibliographystyle{unsrt}

\title[Orthogonal polynomials and the 3nj-Wigner symbols]{Orthogonal
polynomials of several discrete variables and the 3nj-Wigner symbols: applications to spin networks}

\author{M Lorente}
\address {Departamento de F\'{\i}sica, Universidad de Oviedo, 33007 Oviedo, Spain}

\begin{abstract}
The use of orthogonal polynomials for integral models on the lattice is applied to the
3nj-symbols that appear in the coupling of several angular momenta. These symbols are connected to the
Ponzano-Regge method to solve the Einstein equations on a discrete Riemannian manifold.
\end{abstract}


\section{Classical orthogonal polynomials of one discrete variable}

These polynomials satisfy a difference equation of hypergeometric type such that the difference derivatives
of the some polynomial satisfy a similar equation. The discrete variable can be consider of two types. 

\begin{enumerate}
\item[a)] {\it On homogeneous lattice}: $x=0,1,2, \cdots$

The corresponding polynomials satisfy a difference equation
\[\sigma (x)\Delta \nabla p_n(x)+\tau (x)\Delta p_n(x)+\lambda _np_n(x)=0\]

with $\Delta f(x)=f(x+1)-f(x),\quad \nabla f(x)=f(x)-f(x-1)$, $\sigma (x)$ and $\tau (x)$ are functions of
second and first order respectively; 

an orthogonality relation
$$\sum\limits_{x=a}^{b-1} {p_n(x)\,}p_m(x)\,\rho (x)=d_n^2\,\delta _{mn}$$ with $\rho (x)$a weight
function and $d_n$ a normalization constant. To these polynomials correspond the Meixner, Kravchuk,
Charlier and Hahn polynomials [1]
\end{enumerate}

\begin{enumerate}
\item[b)] {\it On non homogeneous lattice}: $x=x(s), \; s=0,1,2, \cdots$ the polynomials satisfy a
difference equation
$$\sigma \left[ {x(s)} \right]{\Delta  \over {\Delta x\left( {s-{1 \over 2}} \right)}}{{\nabla y_n(x)}
\over {\nabla x(s)}}+{1 \over 2}\tau \left[ {x(s)}
\right]\left\{ {{{\Delta y_n(x)} \over {\Delta x(s)}}+{{\nabla y_n(x)} \over {\nabla x(s)}}}
\right\}+\lambda _ny_n(x)=0;$$ 
an orthogonalit relation
$$\sum\limits_{s=a}^{b-1} {y_n(s)}\,y_m(s)\,\rho _n(x)\,\Delta x\left( {s-{1 \over 2}}
\right)=d_n^2\,\delta _{nm}.$$
\end{enumerate}
 The corresponding polynomials are classified according to the latice function:

For $x(s)=s(s+1)$, we have the Racach and dual Hahn polynomials

For $x(s)=q^s$  or ${{q^s-q^{-s}} \over 2}$, we have the $q$-Kravchuck, $q$-Meixner, $q$-Charlier and $q$-Hahn
polynomials.

For $x(s)={{q^s+q^{-s}} \over 2}$ or ${{q^{is}+q^{-is}} \over 2}$, we have the $q$-Racach and $q$-dual Hahn
polynomials [2].

\section{Generalized Clebsch-Gordon coeficients and generalized $3nj$-Wigner symbols}

If two angular momentum operators are coupled to give a total angular momentum $J=J_1+J_2$ the new basis can
be expressed in terms of the old ones

$$\left| {j_1j_2jm} \right\rangle =\sum\limits_{m_1+m_2=m}^{} {\left\langle {{j_1j_2m_1m_2}} \mathrel{\left |
{\vphantom {{j_1j_2m_1m_2} {j_1j_2j_m}}} \right. \kern-\nulldelimiterspace} {{j_1j_2jm}} \right\rangle
}\left| {j_1j_2m_1m_2} \right\rangle .$$
The symmetry properties of the Clebsch-Gordon coefficients in this expansion are more patent if one sbstitutes
them by the Wigner symbols
$$\left\langle {{j_1j_2jm}} \mathrel{\left | {\vphantom {{j_1j_2j_m} {j_1j_2m_1m_2}}} \right.
\kern-\nulldelimiterspace} {{j_1j_2m_1m_2}} \right\rangle =(-1)^{j_1-j_2+j-m}\sqrt {2j+1}\,\left(
{\matrix{{j_1}&{j_2}&j\cr {m_1}&{m_2}&{-m}\cr }} \right)$$
Similarly if we couple three angular momentum operator we obtain a new basis in terms of the old ones:

$$\left| {j_1j_2j_3j_{12}jm} \right\rangle =\sum\limits_{}^{} {\left\langle {{j_1j_2j_3m_1m_2m_3}}
\mathrel{\left | {\vphantom {{j_1j_2j_3m_1m_2m_3} {j_1j_2j_3j_{12}j_m}}} \right. \kern-\nulldelimiterspace}
{{j_1j_2j_3j_{12}jm}} \right\rangle }\left| {j_1j_2j_3m_1m_2m_3} \right\rangle $$
for the coupling $\left( {J_1+J_2} \right)+J_3=J$, 

$$\left| {j_1j_2j_3j_{23}jm} \right\rangle =\sum {\left\langle {{j_1j_2j_3m_1m_2m_3}} \mathrel{\left |
{\vphantom {{j_1j_2j_3m_1m_2m_3} {j_1j_2j_3j_{23}jm}}} \right. \kern-\nulldelimiterspace}
{{j_1j_2j_3j_{23}jm}} \right\rangle }\left| {j_1j_2j_3m_1m_2m_3} \right\rangle $$
for the coupling $J_1+\left( {J_2+J_3} \right)=J$.

Both bases are related by some matrix $U\left( {j_{12},j_{23}} \right)$ that can be written in terms of
generalized $6j$-Wigner symbol

$$U\left( {j_{12},j_{23}} \right)=(-1)^{j_1+j_2+j_3+j}\sqrt {\left( {2j_{12}+1} \right)\left( {2j_{23}+1}
\right)}\left\{ { {\matrix{{j_1}&{j_2}&{j_{12}}\cr {j_3}&j&{j_{23}}\cr }} } \right\}$$
 In similar fashion can be written the generalized Clebsch-Gordon coefficients and generalized
$3nj$-Wigner symbols [3]. The algebraic properties of these symbols can be represented by geometrical graphs
[3].

\section{$3nj$-symbols as orthogonal polynomials of several discrete variable}

The 6j-symbols are proportional to the Racah polynomials through the following relation [2]
$$(-1)^{j_1+j_2+j_{23}}\sqrt {\left( {2j_{12}+1} \right)\left( {2j_{23}+1} \right)}\left\{ {\left(
{\matrix{{j_1}&{j_2}&{j_{12}}\cr {j_3}&j&{j_{23}}\cr }} \right)} \right\}={{\sqrt {\rho (x)}} \over
{d_n}}u_n^{(\alpha ,\beta )}(x,a,b)$$
with 

$x(s)=s(s+1)\;,\;s=j_{23}$

$a=j_3-j_2\quad,\quad b=j+j_3+1\quad,\quad n=j_{12}-j_1+j_2-j$

$\alpha =j_1-j_2-j_3+j\quad ,\quad \beta =j_1-j_2+j_3-j$

Using the assymptotic limit of the Racah polynomials and the connections between the Jacobi polynomials and the
Wigner little functions one can prove the following approximation of the $6j$-sumbols when $j_1\sim
j_2\sim j_3\sim j\gg j_{12}$
\begin{equation}
\left\{ {\matrix{{j_1}&{j_2}&{j_{12}}\cr {j_3}&j&{j_{23}}\cr }} \right\}\simeq
{{(-1)^{j_2+j_3+j_{23}}}
\over {\sqrt {j_1+j_2+1}\sqrt {j_3+j+1}}}d_{j_1-j_2,j_3-j}^{j_{12}}\left( \vartheta  \right)
\end{equation}
with 
$$\cos \theta ={{\left( {2j_{23}+1} \right)^2-\left( {j_1+j_2+1} \right)^2-\left( {j_3+j+1} \right)^2}
\over {2\left( {j_1+j_2+1} \right)\left( {j_3+j+1} \right)}}$$
The $3nj$-symbols of the first and second kind can be written in terms of $6j$-symbols, and therefore in terms
of product of Racah polynomials, giving rise to orthogonal polynomials of several discrete variables. To
illustrate this take, f.i., the $12j$-symbol of the second kind as a combination of $6j$-symbols.

\begin{eqnarray*}
\left\{ {\matrix{{j_1}&{j_2}&{j_3}&{j_4}\cr
{l_1}&{l_2}&{l_3}&{l_4}\cr
{k_1}&{k_2}&{k_3}&{k_4}\cr
}} \right\} &=\sum\limits_x^{} {\left( {2x+1} \right)} 
\left( {-1} \right)^{R_n+4x} 
 \left\{
{\matrix{{j_1}&{k_1}&x\cr {k_2}&{j_2}&{l_1}\cr
}} \right\} \\
& \left\{ {\matrix{{j_2}&{k_2}&x\cr
{k_3}&{j_3}&{l_2}\cr
}} \right\}\left\{ {\matrix{{j_3}&{k_3}&x\cr
{k_4}&{j_4}&{l_3}\cr
}} \right\}\left\{ {\matrix{{j_4}&{k_4}&x\cr
{k_1}&{j_1}&{l_4}\cr
}} \right\}
\end{eqnarray*}

Here $R_n=\sum\limits_{i=1}^{4} {\left(j_i+l_i+k_i\right)}$. Substituting each 6j-symbol for the corresponding
Racah polynomial we obtain:
$$\left\{ {\matrix{{j_1}&{j_2}&{j_3}&{j_4}\cr {l_1}&{l_2}&{l_3}&{l_4}\cr {k_1}&{k_2}&{k_3}&{k_4}\cr }}
\right\}=\sum\limits_x^{} {{1 \over {2x+1}}}\prod\limits_{i=1}^4 {{{\sqrt {\rho \left( {l_i} \right)}} \over
{d_{n_i}}}\;}u_{n_i}^{\left( {\alpha _i,\beta _i} \right)}\left( {l_i} \right)\equiv p_n\left(
{l_1\;l_2\;l_3\;l_4} \right)$$
which is a polynomial of four discrete variables.

For the assymptotic limit we find
$$\left\{ {\matrix{{j_1}&{j_2}&{j_3}&{j_4}\cr {l_1}&{l_2}&{l_3}&{l_4}\cr {k_1}&{k_2}&{k_3}&{k_4}\cr }}
\right\}\approx \sum\limits_x^{} {\left( {2x+1} \right)}\prod\limits_{i=1}^4 {{1 \over
{j_i+k_i+1}}\;}d_{j_i-k_i,j_{i+1}-k_{i+1(\bmod 4)}}^x\left( {\vartheta _i} \right)$$
These formulas can be easily generalized to any $3nj$-symbols of first and second kind.

\section{Application to spin networks and to Ponzano-Regge integral action}

Penrose has proposed a model for the space and time in which the underlying structure is given by a set of
interactions between elementary units that satisfy the coupling of angular momentum operators, called spin
networks [4]. One particular case of these networks can be described by the graphs of $3nj$-symbols. From
different point of view Regge has proposed a method to calculate Einstein action by the approximation of
curved riemannian manifold by a polyedron built up of triangles. Later Ponzano and Regge applied the properties
of $6j$-symbols to calculate the sum action over this triangulation [5]

Let $M$ be a riemannian manifold that is approximated by a polyedron with boundury $D$ and it is decomposed
into $p$ tetrahedra $T_k$ represented by $6j$-symbols.

The polyedron give rise to triangular faces $f$, represented by $3j$-symbols, and to $q$ internal edges $x_i$,
as well as to external ones $l_i$ with respect to the boundary $D$.

Ponzano and Regge define the sum
\begin{equation}
S=\sum\limits_{x_i}^{} {\prod\limits_{k=1}^p {T_k\left( {-1} \right)^\varphi }}\prod\limits_{i=1}^q {\left(
{2x_i+1} \right)}
\end{equation}
When $l_i\to \infty ,\quad \hbar \to 0,\quad \hbar l_i\to \rm finite$ we recovered the continuous
manifold. In order to compute the $6j$-symbols in te classical limit, we uses the assymptotic formula [2]

\newpage

\begin{eqnarray*}
d_{mm'}^j\left( \theta  \right) &\approx \left( {-1} \right)^{m-m'}\sqrt {{2 \over {\pi
\left( {j-m}
\right)}}}\left( {{{2j+m-m'+1} \over {2j-m+m'+1}}} \right)^{{{m+m'} \over 2}}\\
&{{\cos \left[ {\left( {j+{1
\over 2}} \right)\theta -\left( {m-m'+{1 \over 2}} \right){\pi  \over 2}} \right]} \over {\sqrt {\sin \theta
}}}
\end{eqnarray*}
at $m\sim m'\sim 1$, $j>>1$. Substituting this expression in (1) with $m=j_1-j_2\;,\;m'=j_4-j_5\;,\;j=j_6$ and
taking the edges of the tetrahedra $j_1+{1 \over 2},\ldots ,j_6+{1 \over 2}$, very large except $j_6$, we have

\begin{eqnarray}\nonumber\left\{ {\matrix{{j_1}&{j_2}&{j_3}\cr {j_4}&{j_5}&{j_6}\cr }} \right\} &\approx & {1
\over {\sqrt {12 \pi V}}}\cos \left\{\left( {j_6+{1 \over 2}} \right)\theta -\left( {j_1+{1 \over 2}}
\right){\pi \over 2}\; + \right.\\
\nonumber &\quad &\left.+\left(        {j_2+{1 \over 2}} \right){\pi  \over 2} 
 -\left( {j_4+{1 \over 2}} \right){\pi  \over
2}+\left( {j_5+{1 \over 2}} \right){\pi  \over 2}+{\pi  \over 4} \right\}= \\
 &=&{1 \over {\sqrt {12\pi V}}}\cos
\left\{ {\sum\limits_{i=1}^{6} {\left( {j_i+{1 \over 2}} \right)}\theta _i+{\pi  \over 4}}
\right\}
\end{eqnarray}
where $\theta _i$ is the dihedral angle for the edge $j_i$ and $V={1 \over 6}\left( {j_1+{1 \over 2}}
\right)\left( {j_4+{1 \over 2}} \right)\left( {j_6+{1 \over 2}}
\right) \rm sen \; \theta $. Note the formula (3) has been proved rigurously by Roberts [5]. Introducing
formula (3) in formula (2), Ponzano and Regge proved that it leads in the continuous limit to the integral
action of the general relativity.

This work has been partially supported by M.I.C. (grant FM2000-0357) Spain

\vspace{2cm}

e-mail: lorentemiguel@uniovi.es

\end{document}